\documentclass[11pt]{article}
\usepackage{amsmath}
\usepackage{amsfonts}
\usepackage{amssymb}
\usepackage{graphicx}
\usepackage{bm}
\usepackage{color}
\usepackage{amscd}

\def\bea{\begin{eqnarray}}
\def\eea{\end{eqnarray}}

\begin{document}
\begin{center}
\LARGE {\bf Asymptotically spacelike warped anti-de Sitter spacetimes in generalized minimal massive gravity}
\end{center}

\begin{center}
{M. R. Setare \footnote{E-mail: rezakord@ipm.ir}\hspace{1mm} ,
H. Adami \footnote{E-mail: hamed.adami@yahoo.com}\hspace{1.5mm} \\
{\small {\em  Department of Science, University of Kurdistan, Sanandaj, Iran.}}}\\

\end{center}

\begin{center}
{\bf{Abstract}}\\
 In this paper we show that warped AdS$_{3}$ black hole spacetime is a solution of the generalized minimal massive gravity (GMMG) and introduce suitable boundary conditions for asymptotically warped AdS$_{3}$ spacetimes. Then we find the Killing vector fields such that transformations generated by them preserve the considered boundary conditions. We calculate the conserved charges which correspond to the obtained Killing vector fields and show that the algebra of the asymptotic conserved charges is given as the semi direct product of the Virasoro algebra with $U(1)$ current algebra. We use a particular Sugawara construction to reconstruct the conformal algebra. Thus, we are allowed to use the Cardy formula to calculate the entropy of the warped black hole. We demonstrate that the gravitational entropy of the warped black hole exactly coincide with what we obtain via Cardy's formula. As we expect the warped Cardy formula also give us exactly the same result which we obtain from usual Cardy's formula. We calculate mass and angular momentum of the warped black and then check that obtained mass, angular momentum and entropy satisfy first law of the black hole mechanics. According to the results of this paper we belief that the dual theory of the warped AdS$_{3}$ black hole solution of GMMG is a Warped CFT.
\end{center}

\section{Introduction}\label{1.0}
In 1992 Ba$\tilde{\text{n}}$ados, Teitelboim and Zanelli (BTZ) \cite{1',2'} showed that $(2+1)$-dimensional Einstein
gravity in the presence of negative cosmological constant has a black hole solution. This black hole is described by two (gravitational)
parameters, the mass $M$ and the angular momentum (spin) $J$. It is locally AdS and thus
it differs from Schwarzschild and Kerr solutions since it is asymptotically anti-de-Sitter
instead of flat spacetime. Additionally, it has no curvature singularity at the origin. AdS
black holes, are members of this two-parametric family of BTZ black holes and they are
very interesting in the framework of string theory and black hole physics \cite{3',4'}. But we know that the Einstein gravity in three dimensions (with or without cosmological constant) exhibits
no propagating physical degrees of freedom \cite{5',6'}. But adding the gravitational Chern-Simons term produces a
propagating massive graviton \cite{7'}. The resulting theory
is called topologically massive gravity (TMG). In this case the theory exhibits both gravitons and black holes.  Unfortunately there is a problem in this model, with the usual
sign for the gravitational constant, the massive excitations of TMG carry negative energy. In the absence of a cosmological constant, one can
change the sign of the gravitational constant, but if $\Lambda <0$, this will give a negative mass to the BTZ
black hole, so the existence of a stable ground state is in doubt in this model \cite{11}. However 3-dimensional gravity models are rather interesting from the point of view of the AdS/CFT correspondence, in particular, the Minimal Massive Gravity (MMG) model \cite{9'} has attracted some attention as a theory that circumvents the difficulty of defining a bulk theory with positive-energy propagating modes which at the same time has a CFT dual with positive central charges. This has been called the ``bulk-boundary unitarity clash'' by the authors of \cite{9'}. The paper \cite{8} introduces Generalized Minimal Massive Gravity (GMMG), an interesting modification of MMG. As has been shown in \cite{8}, GMMG also avoids the aforementioned ``bulk-boundary unitarity clash''. Hamiltonian analysis show that the GMMG model has no Boulware-Deser ghosts and this model propagate only two physical modes. So these models are viable candidate for semi-classical limit of a unitary quantum $3D$ massive gravity.\\
The realization of the existence of three-dimensional (3D) black holes deepened our understanding of 3D gravity. In this context an important role is played by the notion of asymptotic symmetry. This notion was applied with success some time ago to asymptotically $AdS_3$
spacetimes, to show that the asymptotic symmetry group (ASG) of $AdS_3$ is the conformal
group in two dimensions \cite{19}.
The authors of \cite{21} have introduced spacelike stretched $AdS_3$ as a new vacuum of TMG, which could be a stable ground state of this model \cite{15',23} (see also \cite{18',19',20'}). More than this, another reason for interest to the warped $AdS$ spacetime, is that they emerge
in the near horizon geometry of extremal Kerr black holes and this fact is important in the context of Kerr/CFT \cite{21'}. In the paper \cite{23}, the correctness of the hypothesis formulated in \cite{21} have been investigated. The authors of  \cite{23} have obtained that the asymptotic symmetry of the spacelike stretched $AdS_3$  sector of TMG is a 2-dimensional conformal
symmetry with central charges. Recently warped $AdS_3$ black holes have been studied in generic higher derivatives gravity theories in $2+1$ dimensions \cite{15}. According to earlier investigations on the warped $AdS_3$, the ASG of these spacetimes instead of the Brown-Henneaux conformal symmetry group, is the semi-direct product of a Virasoro algebra and a $U(1)$ affine Kac-Moody algebra \cite{23',24',25',23,16}. It is the symmetry of warped conformal field theory (WCFT) in 2-dimension. The authors of \cite{15} have reproduced the match between the Bekenstein-Hawking and WCFT entropies in the case of new massive gravity (NMG) \cite{27'}. \\
For systems that admit 2D CFTs as duals, the Cardy formula \cite{24} can be applied
directly. This formula gives the entropy of a CFT in terms of the central charge $c$ and
the eigenvalue of the Virasoro operator $l_0$. However, it should be pointed out that this
evaluation is possible as soon as one has explicitly shown (e.g using the $AdS_d/CFT_{d-1}$
correspondence) that the system under consideration is in correspondence with a 2D
CFT \cite{12',13'}.\\
In this paper we consider the spacelike warped $AdS_3$ black hole solutions of GMMG.
The paper is organized as follows. In section 2 we briefly review the quasi-local conserved charges in  Chern-Simons-like theories of gravity (CSLTGs). Since GMMG is an example of CSLTGs, we present generic formula for entropy of the black hole solutions of such theories, which previously has obtained in \cite{7}. Then in section 3 we introduce the GMMG model and present its equations of motion. In section 4 we show that the spacelike warped $AdS_3$ black hole is a solution of GMMG model. In section 5 following paper \cite{16} we introduce the appropriate boundary conditions for asymptotically spacelike warped AdS$_{3}$ spacetimes. We show that the algebra among the asymptotic Killing vectors is the semi direct product of the Witt algebra with the $U(1)$ current algebra. In section 6 we find the conserved charges for asymptotically spacelike warped AdS$_{3}$ spacetime solutions of GMMG. In section 7 we, at first we obtain the conserved charge corresponds to the asymptotic Killing vector, then we find the algebra of conserved charges. After that we show that the algebra of asymptotic conserved charges for the spacelike warped AdS$_{3}$ black hole is semi direct product of the Virasoro algebra with $U(1)$ current algebra. In section 8 we use a particular Sugawara construction \cite{22}
to reconstruct the conformal algebra. We then show that the Cardy formula
reproduces exactly the Bekenstein-Hawking entropy of the spacelike stretched $AdS_3$ black hole solution of GMMG. We show that this entropy can be reproduce by warped version of the Cardy formula also. In this section we obtain the mass and angular momentum of black hole, and show that these quantities satisfy the first law of black hole mechanics. In section 9 we obtain the entropy of black hole using our generic entropy formula which we presented in section 2. Last section contain our conclusions.

\section{Quasi-local conserved charges in Chern-Simons-like theories of gravity}\label{2.0}
The Lagrangian 3-form of the Chern-Simons-like theories of gravity is given by \cite{1}
\begin{equation}\label{1}
  L=\frac{1}{2} \tilde{g}_{rs} a^{r} \cdot da^{s}+\frac{1}{6} \tilde{f}_{rst} a^{r} \cdot a^{s} \times a^{t}.
\end{equation}
In the above Lagrangian $ a^{ra}=a^{ra}_{\hspace{3 mm} \mu} dx^{\mu} $ are Lorentz vector
 valued one-forms where, $r$ and $a$ indices refer to flavour and Lorentz indices, respectively.
 We should mention that, here, the wedge products of Lorentz-vector valued one-form fields are implicit.
 Also, $\tilde{g}_{rs}$ is a symmetric constant metric on the flavour space and $\tilde{f}_{rst}$
 is a totally symmetric "flavour tensor" which are interpreted as the coupling constants.
 We use a 3D-vector algebra notation for Lorentz vectors in which contractions
with $\eta _{ab}$ and $\varepsilon ^{abc}$ are denoted by dots and crosses,
 respectively \footnote{Here we consider the notation used in \cite{1}.}.
 It is worth saying that $a^{ra}$ is a collection of the dreibein $e^{a}$, the dualized
 spin-connection $\omega ^{a}$, the auxiliary field
 $ h^{a}_{\hspace{1.5 mm} \mu} = e^{a}_{\hspace{1.5 mm} \nu} h^{\nu}_{\hspace{1.5 mm} \mu} $ and
 so on \footnote{That is $a^{r}=\{ e,\omega, h, \cdots \}$, for instance, for
 $r=e$ and $r=\omega$ we have $a^{e}=e$ and $a^{\omega}=\omega$.}. Also for all interesting CSLTG we have $\tilde{f}_{\omega rs} = \tilde{g}_{rs}$ \cite{2}.\\
The total variation of $a^{ra}$ induced by a diffeomorphism generator $\xi$ is \cite{4}
\begin{equation}\label{2}
  \delta _{\xi} a^{ra} = \mathfrak{L}_{\xi} a^{ra} -\delta ^{r} _{\omega} d \chi _{\xi} ^{a} ,
\end{equation}
where $\chi _{\xi} ^{a}= \frac{1}{2} \varepsilon ^{a} _{\hspace{1.5 mm} bc} \lambda _{\xi}^{bc} $ and $\lambda _{\xi}^{bc}$
 is generator of the Lorentz gauge transformations $SO(2, 1)$. Also, $ \delta ^{r} _{s} $
  denotes the ordinary Kronecker delta and the Lorentz-Lie derivative along a vector field $\xi$ is
 denoted by $\mathfrak{L}_{\xi}$. The Lorentz-Lie derivative of a Lorentz tensor-valued
 $p$-form $\mathcal{A}^{a \cdots}_{b \cdots}$ is defined by
\begin{equation}\label{90}
  \mathfrak{L}_{\xi} \mathcal{A}^{a \cdots}_{b \cdots}= \pounds_{\xi} \mathcal{A}^{a \cdots}_{b \cdots} + \lambda^{\hspace{1 mm} a}_{\xi \hspace{1 mm} c} \mathcal{A}^{c \cdots}_{b \cdots}+ \cdots - \lambda^{\hspace{1 mm} c}_{\xi \hspace{1 mm} b}\mathcal{A}^{a \cdots}_{c \cdots} - \cdots.
\end{equation}
Dreibein and spin-connection \footnote{Spin-connection $\omega^{ab}$ and
 dualized spin-connection $\omega^{a}$ are related as $\omega^{a}= \frac{1}{2}\varepsilon^{abc}\omega_{bc}$.} transform like $e \rightarrow \Lambda e$ and $\omega \rightarrow \Lambda \omega \Lambda^{-1} + \Lambda d \Lambda^{-1}$ under Lorentz gauge transformations, where $\Lambda= \exp (\lambda) \in SO(2,1)$. Total variation induced by a vector field $\xi$ is a combination of variations due to a diffeomorphism and an infinitesimal Lorentz gauge transformation \cite{7}. It is obvious that the total variation of a Lorentz tensor-valued $p$-form is equal to its Lorentz-Lie derivative and the extra term in total variation of $a^{ra}$ comes from the transformation law
of spin-connection under Lorentz gauge transformations. Total variation of $a^{ra}$ is covariant under Lorentz gauge transformations as well as diffeomorphism. Hence we are allowed to use covariant phase space method to obtain conserved charges in CSLTG. The arbitrary variation of the Lagrangian \eqref{1} is
\begin{equation}\label{91}
  \delta L = \delta a^{r} \cdot E_{r} + d \Theta (a, \delta a),
\end{equation}
with
\begin{equation}\label{92}
   E_{r}^{\hspace{1.5 mm} a} = \tilde{g}_{rs} d a^{sa} + \frac{1}{2} \tilde{f}_{rst} (a^{s} \times a^{t})^{a} , \hspace{0.7 cm} \Theta (a, \delta a) = \frac{1}{2} \tilde{g}_{rs} \delta a^{r} \cdot a^{s}.
\end{equation}
where $ E_{r}^{\hspace{1.5 mm} a} =0$ are the equations of motion and $\Theta (a, \delta a)$ is surface term.
 The total variation of the Lagrangian induced by diffeomorphism generator $\xi$ can be written as
 \footnote{$i_{\xi}$ denotes interior product in $\xi$.}
\begin{equation}\label{93}
  \delta_{\xi} L = \mathfrak{L}_{\xi} L + d \psi_{\xi}=d \left( i_{\xi} L + \psi_{\xi} \right),
\end{equation}
with $\psi _{\xi} = \frac{1}{2} \tilde{g}_{\omega r} d \chi_{\xi} \cdot a^{r}$. Also, the total variation of the surface term is
\begin{equation}\label{94}
  \delta_{\xi} \Theta (a, \delta a) = \mathfrak{L}_{\xi} \Theta (a, \delta a) + \Pi_{\xi},
\end{equation}
with $\Pi _{\xi}=\frac{1}{2} \tilde{g}_{\omega r} d \chi_{\xi} \cdot \delta a^{r}$. Now we assume that the variation of Lagrangian \eqref{1} is generated by a vector field $\xi$. For generality, we assume that vector field $\xi$ depends on dynamical fields. By using the Bianchi identities, we find off-shell Noether current
\begin{equation}\label{95}
  J_{\xi}= \Theta (a, \delta_{\xi} a) - i_{\xi} L - \psi_{\xi} + i_{\xi} a^{r} \cdot E_{r} - \chi _{\xi} \cdot E_{\omega}
\end{equation}
which is conserved off-shell, i.e. we have $dJ_{\xi}=0$ off-shell.
 By virtue of the Poincare lemma, one can obtain off-shell Noether charge density
 $K_{\xi}$ so that $J_{\xi}= d K_{\xi}$. By taking variation from Eq.\eqref{95}
 with respect to dynamical fields
\footnote{We denote variation with respect to dynamical fields by $\hat{\delta}$.} and by making some calculations
 one can define extended off-shell ADT
 \footnote{ ADT stands for Abbott, Deser and Tekin.} current as \cite{3}
\begin{equation}\label{96}
\begin{split}
   \mathcal{J}_{\text{ADT}} (a , \delta a, \delta _{\xi} a)=  & \hat{\delta} a^{r} \cdot i_{\xi} E_{r} + i_{\xi} a^{r} \cdot \hat{\delta} E_{r} - \chi _{\xi} \cdot \hat{\delta} E_{\omega} \\
     & + \tilde{g} _{rs} \delta _{\xi} a^{r} \cdot \hat{\delta} a^{s}.
\end{split}
\end{equation}
If we assume that $\xi$ is a Killing vector field then the last term in Eq.\eqref{96} vanishes and extended off-shell ADT current reduces to the generalized off-shell ADT current \cite{4}. Also, if we consider on-shell case, i.e. $E_{r}=\delta E_{r}=0$, then extended off-shell ADT current reduces to symplectic current \cite{31}. The current \eqref{96} is conserved off-shell, i.e. $d\mathcal{J}_{\text{ADT}}=0$, so by virtue of the Poincare lemma, one can find corresponding extended off-shell ADT charge so that $\mathcal{J}_{\text{ADT}}=d \mathcal{Q}_{\text{ADT}}$. Therefore we can define quasi-local conserved charge perturbation associated with a field dependent vector field $\xi$ as
\begin{equation}\label{3}
\begin{split}
   \hat{\delta} Q ( \xi )  & =\frac{1}{8 \pi G} \int_{\Sigma} \mathcal{Q}_{\text{ADT}} \\
     & =\frac{1}{8 \pi G} \int_{\Sigma} \left( \tilde{g}_{rs} i_{\xi} a^{r} - \tilde{g} _{\omega s} \chi _{\xi} \right) \cdot \hat{\delta} a^{s},
\end{split}
\end{equation}
where $G$ denotes the Newtonian gravitational constant and $\Sigma$ is a spacelike codimension two surface. We can take an integration from \eqref{3} over the one-parameter path on the solution space \cite{5,6} and then we find that
\begin{equation}\label{4}
  Q ( \xi )  = \frac{1}{8 \pi G} \int_{0}^{1} ds \int_{\Sigma} \left( \tilde{g}_{rs} i_{\xi} a^{r} - \tilde{g} _{\omega s} \chi _{\xi} \right) \cdot \hat{\delta} a^{s},
\end{equation}
Also, we argued that the quasi-local conserved charge \eqref{4} is not only conserved for the Killing vectors which
 are admitted by spacetime everywhere but also it is conserved for the asymptotically Killing vectors.\\
The entropy of black holes is the conserved charge associated with the
 horizon-generating Killing vector field evaluated at the bifurcation surface \cite{31}.
 Let $\zeta$ denotes horizon-generating Killing vector field.
 The horizon-generating Killing vector field $\zeta$ vanishes on the bifurcation surface $\mathcal{B}$.
 Now, take $\Sigma$ in Eq.\eqref{4} to be the bifurcation surface $\mathcal{B}$ then we have
\begin{equation}\label{97}
 Q ( \zeta )  = -\frac{1}{8 \pi G} \tilde{g} _{\omega r} \int_{\mathcal{B}}  \chi _{\zeta} \cdot a^{r}.
\end{equation}
To obtain an explicit expression for $\chi _{\xi}$, in an appropriate manner, the authors in \cite{17} demand that it must be chosen so that the Lorentz-Lie derivative of dreibein vanishes when $\xi$ is a Killing vector field and then $\chi _{\xi}$ should be provided as follows \cite{7}:
\begin{equation}\label{98}
  \chi _{\xi} ^{a} = i_{\xi} \omega ^{a} + \frac{1}{2} \varepsilon ^{a}_{\hspace{1.5 mm} bc} e^{\nu b} (i_{\xi} T^{c})_{\nu} + \frac{1}{2} \varepsilon ^{a}_{\hspace{1.5 mm} bc} e^{b \mu} e^{c \nu} \nabla _{\mu} \xi _{\nu} .
\end{equation}
Thus, on the bifurcation surface we will have $ \chi _{\zeta}^{a}|_{\mathcal{B}}= \kappa N^{a}$ with $N^{a}=\frac{1}{2} \varepsilon^{abc}n_{bc}$, where $\kappa$ and $n_{ab}$ are respectively surface gravity and bi-normal to the bifurcation surface. Therefore black hole entropy in the CSLTG  can be defined as
\begin{equation}\label{99}
  \mathcal{S}= \frac{2\pi}{\kappa} Q ( \zeta ) = \frac{1}{4G} \tilde{g} _{\omega r} \int_{\mathcal{B}}  N \cdot a^{r}.
\end{equation}
Since the non-zero components of bi-normal $n_{\mu \nu}$ to stationary black hole horizon are $n_{01}= -n_{10}$ and it is normalized to $-2$ so Eq.\eqref{99} reduces to \cite{7}
\begin{equation}\label{5}
  \mathcal{S}=- \frac{1}{4G} \int_{\mathcal{B}} \frac{d \phi}{\sqrt{g_{\phi \phi}}} \tilde{g} _{\omega r} a^{r}_{\phi \phi},
\end{equation}
where $\phi$ is angular coordinate and $g_{\phi \phi}$ denotes the $\phi \text{-} \phi$ component of
 spacetime metric $g_{\mu \nu}$.
We use the above generic entropy formula to obtain the Bekenstein-Hawking entropy of the spacelike
 warped $AdS_3$ black hole solution of GMMG model.
\section{Generalized Minimal Massive Gravity}\label{3.0}
Generalized minimal massive gravity (GMMG) is an example of the Chern-Simons-like theories of gravity \cite{8}. In the GMMG, there are four flavours of one-form, $a^{r}= \{ e, \omega , h, f \}$, and the non-zero components of the flavour metric and the flavour tensor are
\begin{equation}\label{6}
\begin{split}
     & \tilde{g}_{e \omega}=-\sigma, \hspace{1 cm} \tilde{g}_{e h}=1, \hspace{1 cm} \tilde{g}_{\omega f}=-\frac{1}{m^{2}}, \hspace{1 cm} \tilde{g}_{\omega \omega}=\frac{1}{\mu}, \\
     & \tilde{f}_{e \omega \omega}=-\sigma, \hspace{1 cm} \tilde{f}_{e h \omega}=1, \hspace{1 cm} \tilde{f}_{f \omega \omega}=-\frac{1}{m^{2}}, \hspace{1 cm} \tilde{f}_{\omega \omega \omega}=\frac{1}{\mu},\\
     & \tilde{f} _{eff}= -\frac{1}{m^{2}}, \hspace{1 cm} \tilde{f}_{eee}=\Lambda_{0},\hspace{1 cm} \tilde{f}_{ehh}= \alpha .
\end{split}
\end{equation}
where $\sigma$, $\Lambda _{0}$, $\mu$, $m$ and $\alpha$ are a sign, cosmological parameter with dimension of mass squared, mass parameter of Lorentz Chern-Simons term, mass parameter of New Massive Gravity term and a dimensionless parameter, respectively. The equations of motion of  GMMG are \cite{8} (see also \cite{9})
\begin{equation}\label{7}
   - \sigma R (\Omega) + (1 + \sigma \alpha ) D(\Omega) h - \frac{1}{2} \alpha (1 + \sigma \alpha ) h \times h + \frac{\Lambda _{0}}{2} e \times e - \frac{1}{2 m^{2}} f \times f  =0,
\end{equation}
\begin{equation}\label{8}
  - e \times f + \mu (1 + \sigma \alpha ) e \times h - \frac{\mu}{m^{2}} D(\Omega) f + \frac{\mu \alpha}{m^{2}} h \times f=0 ,
\end{equation}
\begin{equation}\label{9}
  R(\Omega) - \alpha D(\Omega) h + \frac{1}{2} \alpha ^{2} h \times h + e \times f =0,
\end{equation}
\begin{equation}\label{10}
  T(\Omega) = 0 ,
\end{equation}
where
\begin{equation}\label{11}
  \Omega = \omega - \alpha h
\end{equation}
is ordinary torsion-free dualized spin-connection. Also, $R(\Omega) = d \Omega + \frac{1}{2} \Omega \times \Omega$ is curvature 2-form, $T(\Omega)= D(\Omega)e$ is torsion 2-form, and $ D(\Omega) $ denotes exterior covariant derivative with respect to torsion-free dualized spin-connection.
\section{Warped black holes as solutions of GMMG}\label{4.0}
Now we consider the stationary black hole metric in ADM form
\begin{equation}\label{12}
  \frac{ds^{2}}{l^{2}}= - N(r)^{2} dt^{2} + \frac{dr^{2}}{4 N(r)^{2} R(r)^{2} } + R(r)^{2} \left( d \phi + N^{\phi}(r) dt\right)^{2},
\end{equation}
where $t$, $r$, $\phi$ and $l$ are time-coordinate, radial-coordinate, angular-coordinate and AdS$_{3}$ space radius, respectively. For the spacelike warped AdS$_{3}$ black hole we have \cite{10}
\begin{equation}\label{13}
  \begin{split}
     R(r)^{2} = & \frac{1}{4} \zeta ^{2} r \left[ \left( 1 - \nu ^{2} \right) r + \nu ^{2} \left( r_{+} + r_{-} \right) + 2 \nu \sqrt{r_{+} r_{-}} \right], \\
      N(r)^{2}= & \zeta^{2} \nu^{2} \frac{\left( r - r_{+} \right)\left( r - r_{-} \right)}{4 R(r)^{2}}, \\
      N^{\phi}(r)= &\left| \zeta \right| \frac{r+\nu \sqrt{r_{+} r_{-}}}{2 R(r)^{2}},
  \end{split}
\end{equation}
where $r_{+}$ and $r_{-}$ are outer and inner horizons, respectively. The parameters appeared in Eq.\eqref{13}, $\zeta$ and $\nu$, allow us to keep contact with \cite{11,12,13,14}\footnote{For further discussion, see \cite{10}.}. The spacetime described by metric \eqref{12} with \eqref{13} admits $SL(2,\mathbb{R})\times U(1)$ as isometry group. Therefore, one can write a symmetric-two tensor $S_{\mu \nu}$ as \cite{15}
\begin{equation}\label{14}
  S_{\mu \nu}=a_{1} g_{\mu \nu} +a_{2} J_{\mu} J_{\nu},
\end{equation}
with
\begin{equation}\label{15}
  J=J^{\mu}\partial_{\mu}=\partial_{t}.
\end{equation}
It should be noted that $J_{\mu} J^{\mu}=l^{2}$ and
\begin{equation}\label{16}
  \nabla _{\mu} J_{\nu} = \frac{\left| \zeta \right|}{2l} \epsilon_{\mu \nu \lambda} J^{\lambda}
\end{equation}
where $\nabla _{\mu}$ is covariant derivative with respect to the Christoffel connection and $  \epsilon_{\mu \nu \lambda} = \sqrt{-g}  \varepsilon_{\mu \nu \lambda}$. Hence, the Ricci tensor can be written as
\begin{equation}\label{17}
  \mathcal{R}_{\mu \nu} = \frac{\zeta^{2}}{2l^{2}}\left( 1-2 \nu ^{2} \right) g_{\mu \nu} - \frac{\zeta^{2}}{l^{4}} \left( 1- \nu ^{2} \right) J_{\mu} J_{\nu},
\end{equation}
and the Ricci scalar is given by
\begin{equation}\label{18}
  \mathcal{R}= \frac{\zeta^{2}}{2l^{2}}\left( 1-4 \nu ^{2} \right).
\end{equation}
It is easy to see that dreibein correspond to the metric \eqref{12} can be written as
\begin{equation}\label{19}
  \begin{split}
     e^{0} = & l N(r) dt, \\
     e^{1} = & \frac{l}{2 R(r) N(r)}dr, \\
     e^{2} = & l R(r) \left( d \phi + N^{\phi} dt\right).
  \end{split}
\end{equation}
To show that the metric \eqref{12} with \eqref{13} is a solution of GMMG we consider following ansatz for $h$ and $f$
\begin{equation}\label{20}
  \begin{split}
     h^{a}_{\hspace{1.5 mm \mu}}= & H_{1} e^{a}_{\hspace{1.5 mm \mu}} +H_{2} J^{a}J_{\mu}, \\
     f^{a}_{\hspace{1.5 mm \mu}}= & F_{1} e^{a}_{\hspace{1.5 mm \mu}} +F_{2} J^{a}J_{\mu},
  \end{split}
\end{equation}
where $H_{1}$, $H_{2}$, $F_{1}$, $F_{2}$ are constant parameters and $J^{a} = e^{a}_{\hspace{1.5 mm \mu}} J^{\mu} $. One can use equations \eqref{15}-\eqref{20}, to show that equations of motion of GMMG \eqref{7}-\eqref{10} reduce to the following equations
\begin{equation}\label{21}
  \frac{\zeta^{2}}{4l^{2}} - \frac{1}{2} \alpha l \left| \zeta \right| H_{2} - \alpha ^{2} H_{1} \left( H_{1} + l^{2} H_{2} \right) - \left( 2F_{1}+l^{2} F_{2} \right)=0,
\end{equation}
\begin{equation}\label{22}
  -\frac{\zeta^{2}}{l^{4}} \left( 1-\nu^{2} \right) + \frac{3 \alpha}{2l} \left| \zeta \right| H_{2} + \alpha ^{2} H_{1} H_{2} +F_{2}=0,
\end{equation}
\begin{equation}\label{23}
  \begin{split}
     & \frac{1}{\mu} \left( 2F_{1}+l^{2} F_{2} \right)-\left( 1+ \alpha \sigma\right) \left( 2 H_{1} + l^{2} H_{2} \right) - \frac{l}{2 m^{2}} \left| \zeta \right| F_{2}  \\
       & - \frac{\alpha}{m^{2}} \left[ 2 H_{1}F_{1} + l^{2} \left( H_{1}F_{2}+H_{2}F_{1} \right) \right]=0,
  \end{split}
\end{equation}
\begin{equation}\label{24}
  -\frac{1}{\mu} F_{2} + \left( 1+ \alpha \sigma\right) H_{2} + \frac{3}{2lm^{2}} \left| \zeta \right| F_{2} +\frac{\alpha}{m^{2}} \left( H_{1}F_{2}+H_{2}F_{1} \right) =0,
\end{equation}
\begin{equation}\label{25}
  \begin{split}
     &- \frac{\zeta^{2}}{4l^{2}} \sigma + \frac{1}{2} \left( 1+ \alpha \sigma\right) l \left| \zeta \right| H_{2} + \alpha \left( 1+ \alpha \sigma\right) H_{1} \left( H_{1} + l^{2} H_{2} \right) \\
       & - \Lambda _{0} + \frac{1}{m^{2}} F_{1} \left( F_{1}+l^{2} F_{2} \right)=0,
  \end{split}
\end{equation}
\begin{equation}\label{26}
  \frac{\zeta^{2}}{l^{4}} \left( 1-\nu^{2} \right) \sigma -\frac{3}{2l} \left( 1+ \alpha \sigma\right) \left| \zeta \right| H_{2}- \alpha \left( 1+ \alpha \sigma\right) H_{1}H_{2} - \frac{1}{m^{2}} F_{1} F_{2}=0.
\end{equation}
Thus, the metric \eqref{12} with \eqref{13} is a solution of GMMG if the introduced parameters satisfy equations \eqref{21}-\eqref{26}.
\section{Asymptotically spacelike warped AdS$_{3}$ spacetimes}\label{5.0}
In this section we follow the paper \cite{16} to introduce appropriate boundary conditions. In this way, for asymptotically spacelike warped AdS$_{3}$ spacetimes, we introduce following boundary conditions
\begin{equation}\label{27}
  \begin{split}
       & g_{tt}=l^{2}+\mathcal{O}(r^{-3}), \hspace{0.7 cm} g_{tr}= \mathcal{O}(r^{-3}), \hspace{0.7 cm} g_{r\phi}= \mathcal{O}(r^{-2}), \\
       & g_{t\phi}=\frac{1}{2} l^{2} \left| \zeta \right| \left[ r + A_{t\phi} (\phi) + \frac{1}{r} B_{t\phi} (\phi) \right] + \mathcal{O}(r^{-2}), \\
       & g_{rr}= \frac{l^{2}}{\zeta^{2} \nu^{2}} \left[ \frac{1}{r^{2}} + \frac{1}{r^{3}} A_{rr} (\phi) + \frac{1}{r^{4}} B_{rr} (\phi) \right]+\mathcal{O}(r^{-5}), \\
       & g_{\phi \phi}= \frac{1}{4} l^{2} \zeta^{2} \left[ \left( 1-\nu^{2} \right) r^{2} + r A_{\phi \phi} (\phi) + B_{\phi \phi} (\phi) \right] + \mathcal{O}(r^{-1}),
  \end{split}
\end{equation}
which are consistent with the metric \eqref{12}. The corresponding components of dreibein are
\begin{equation}\label{28}
\begin{split}
     e^{0}_{\hspace{1.5 mm}t} = & \frac{l \nu}{\sqrt{1-\nu^{2}}} - \frac{l \left[ 2 \left( \nu^{2}-1\right) A_{t\phi}+A_{\phi\phi}\right]}{2r\nu \left( 1 - \nu^{2} \right)^{\frac{3}{2}}} \\
     & + \frac{l}{8r^{2} \nu^{3} \left( 1 - \nu^{2} \right)^{\frac{5}{2}}} \biggl[ 4 A_{t\phi}^{2} \left(\nu^{2}-1 \right)^{3} + A_{\phi\phi}^{2} \left( 4 \nu^{2}-1\right) \\
     & + 4 A_{t\phi} A_{\phi\phi} \left( \nu^{2}-1\right)\left( 2 \nu^{2}-1 \right)+8B_{t\phi}\nu^{2} \left(\nu^{2}-1 \right)^{2}\\
     & +4B_{\phi\phi} \nu^{2} \left(\nu^{2}-1 \right)\biggr]+\mathcal{O}(r^{-3})
\end{split}
\end{equation}
\begin{equation}\label{29}
   e^{1}_{\hspace{1.5 mm} r}= \frac{l}{\zeta \nu r}+\frac{lA_{rr}}{2 \zeta \nu r^{2}}-\frac{l}{8\zeta \nu r^{3}} \left[A_{rr}^{2}-4B_{rr}\right]+\mathcal{O}(r^{-4})
\end{equation}
\begin{equation}\label{30}
  \begin{split}
   e^{2}_{\hspace{1.5 mm} t}= & \frac{l}{\sqrt{1-\nu^{2}}} - \frac{l \left[ 2 \left( \nu^{2}-1\right) A_{t\phi}+A_{\phi\phi}\right]}{2r \left( 1 - \nu^{2} \right)^{\frac{3}{2}}} \\
     & + \frac{l}{8r^{2} \left( 1 - \nu^{2} \right)^{\frac{5}{2}}} \biggl[8B_{t\phi} \left( \nu^{2}-1\right)^{2}+4 A_{t\phi} A_{\phi\phi} \left( \nu^{2}-1\right)\\
     & +4B_{\phi\phi} \left( \nu^{2}-1\right)+3A_{\phi\phi}^{2}\biggr]+\mathcal{O}(r^{-3})
\end{split}
\end{equation}
\begin{equation}\label{31}
  \begin{split}
     e^{2}_{\hspace{1.5 mm} \phi}= & \frac{1}{2} rl\left| \zeta \right| \sqrt{1-\nu^{2}} + \frac{l\left| \zeta \right| A_{\phi\phi}}{4\sqrt{1-\nu^{2}}} \\
       & -\frac{l\left| \zeta \right|}{16r \left( 1 - \nu^{2} \right)^{\frac{3}{2}}} \left[ 4B_{\phi\phi} \left(\nu^{2}-1\right)+A_{\phi\phi}^{2} \right]+\mathcal{O}(r^{-2}),
  \end{split}
\end{equation}
and the rest of them are of the order of $r^{-4}$, that is $\mathcal{O}(r^{-4})$.\\
The metric, under transformation generated by vector field $\xi$, transforms as $\delta _{\xi} g_{\mu \nu}= \pounds_{\xi}g_{\mu \nu} $\footnote{Where $\pounds_{\xi}$ denotes usual Lie derivative along $\xi$.}. The variation generated by the following Killing vector field preserves the boundary conditions \eqref{27}
\begin{equation}\label{32}
  \begin{split}
     \xi^{t}(T,Y)= & T(\phi) - \frac{2 \partial_{\phi}^{2} Y(\phi)}{\left| \zeta \right|^{3} \nu^{4} r}+\mathcal{O}(r^{-2}), \\
     \xi^{r}(T,Y)= & -r \partial_{\phi} Y(\phi)+ \mathcal{O}(r^{-2}),\\
     \xi^{\phi}(T,Y)= & Y(\phi) + \frac{2 \partial_{\phi}^{2} Y(\phi)}{ \zeta ^{4} \nu^{4} r^{2}}+\mathcal{O}(r^{-3}),
  \end{split}
\end{equation}
where $T(\phi)$ and $Y(\phi)$ are two arbitrary periodic functions. The asymptotic Killing vectors \eqref{32} are closed in the Lie bracket and we have
\begin{equation}\label{33}
  \left[ \xi(T_{1},Y_{1}),\xi(T_{2},Y_{2}) \right]=\xi(T_{12},Y_{12}),
\end{equation}
where
\begin{equation}\label{34}
  \begin{split}
     T_{12}(\phi)= & Y_{1}(\phi)\partial_{\phi} T_{2}(\phi)- Y_{2}(\phi)\partial_{\phi} T_{1}(\phi),\\
     Y_{12}(\phi)=  & Y_{1}(\phi)\partial_{\phi} Y_{2}(\phi)-Y_{2}(\phi)\partial_{\phi} Y_{1}(\phi).
  \end{split}
\end{equation}
By introducing Fourier modes $u_{m}=\xi(e^{im\phi},0)$ and $v_{m}=\xi(0,e^{im\phi})$, one can find that
\begin{equation}\label{35}
  \begin{split}
       & \left[ u_{m},u_{n} \right]=0, \\
       &  \left[ v_{m},u_{n} \right]=-n u_{m+n},\\
       & \left[ v_{m},v_{n} \right]=(m-n)v_{m+n}.
  \end{split}
\end{equation}
Therefore the algebra among the asymptotic Killing vectors is the semi direct product of the Witt algebra with the $U(1)$ current algebra. Under the action of a generic asymptotic symmetry generator $\xi$ spanned by \eqref{32}, the dynamical fields transform as
\begin{equation}\label{36}
  \begin{split}
     \delta _{\xi} A_{t\phi} = & \partial_{\phi} \left[ Y(\phi) A_{t\phi}(\phi) \right]+\frac{2}{\left| \zeta \right|} \partial_{\phi}T(\phi), \\
     \delta _{\xi} A_{rr} =  &  \partial_{\phi} \left[ Y(\phi) A_{rr}(\phi) \right], \\
     \delta _{\xi} A_{\phi\phi} =  &  \partial_{\phi} \left[ Y(\phi) A_{\phi\phi}(\phi) \right] +\frac{4}{\left| \zeta \right|} \partial_{\phi}T(\phi),
  \end{split}
\end{equation}
\begin{equation}\label{37}
\begin{split}
    \delta _{\xi} B_{t\phi} = & Y(\phi) \partial_{\phi} B_{t\phi}(\phi) + 2 B_{t\phi}(\phi) \partial_{\phi} Y(\phi) - \frac{2}{\zeta^{4} \nu^{4}} \partial_{\phi}^{3} Y(\phi) ,\\
    \delta _{\xi} B_{rr} = & Y(\phi) \partial_{\phi} B_{rr}(\phi) + 2 B_{rr}(\phi) \partial_{\phi} Y(\phi) ,\\
    \delta _{\xi} B_{\phi\phi} = & Y(\phi) \partial_{\phi} B_{\phi\phi}(\phi) + 2 B_{\phi\phi}(\phi) \partial_{\phi} Y(\phi)+\frac{4}{\left| \zeta \right|} A_{t\phi}(\phi) \partial_{\phi} T(\phi) \\
     & - \frac{4\left( 1+\nu^{2} \right)}{\zeta^{4} \nu^{4}} \partial_{\phi}^{3} Y(\phi).
\end{split}
\end{equation}
We are interested in solutions which are asymptotically spacelike warped AdS$_{3}$. Thus, we demand that equations \eqref{16} and \eqref{17} hold asymptotically, i.e.
\begin{equation}\label{38}
  \nabla _{\mu} J_{\nu} - \frac{\left| \zeta \right|}{2l} \epsilon_{\mu \nu \lambda} J^{\lambda}=\mathcal{O}(r^{-1}),
\end{equation}
\begin{equation}\label{39}
  \mathcal{R}_{\mu \nu} - \frac{\zeta^{2}}{2l^{2}}\left( 1-2 \nu ^{2} \right) g_{\mu \nu} + \frac{\zeta^{2}}{l^{4}} \left( 1- \nu ^{2} \right) J_{\mu} J_{\nu}=\mathcal{O}(r^{-1}).
\end{equation}
By substituting Eq.\eqref{27} into the equations \eqref{38} and \eqref{39}, we find that
\begin{equation}\label{40}
  \begin{split}
       & A_{\phi\phi}(\phi)=\nu^{2} A_{rr}(\phi)+2A_{t\phi}(\phi),  \\
       & B_{\phi\phi}(\phi)=\nu^{2} \left[ B_{rr}(\phi)+2B_{t\phi}(\phi)-A_{rr}(\phi)^{2} \right]+A_{t\phi}(\phi)^{2}+2B_{t\phi}(\phi).
  \end{split}
\end{equation}
Hence, the metric \eqref{27} solves equations of motion of GMMG asymptotically when equations \eqref{21}-\eqref{26} and \eqref{40} are satisfied. We emphasis, following above analysis, that equations \eqref{16}, \eqref{17} and \eqref{20} are held asymptotically.
\section{Conserved charges of asymptotically spacelike warped AdS$_{3}$ spacetimes in GMMG}\label{6.0}
We first want to simplify the expression for conserved charge perturbation \eqref{3} for asymptotically spacelike warped AdS$_{3}$ spacetimes \eqref{27} in the context of GMMG. To this end, we use equations \eqref{20}-\eqref{26}, \eqref{6}, \eqref{11} and \eqref{17}. After some calculations we find that
\begin{equation}\label{41}
  \begin{split}
      \hat{\delta} Q(\xi) = \frac{1}{8\pi G} \int_{\Sigma}\biggl\{& - \left( \sigma + \frac{\alpha H_{1}}{\mu} + \frac{F_{1}}{m^{2}}\right) \left[ i_{\xi} e \cdot\hat{\delta} \Omega + \left( i_{\xi} \Omega - \chi _{\xi}\right)\cdot \hat{\delta} e \right]  \\
       & +\frac{1}{\mu} \left( i_{\xi} \Omega - \chi _{\xi}\right) \cdot \hat{\delta} \Omega + \alpha H_{2} \left( \frac{\alpha H_{2}}{\mu} + \frac{2 F_{2}}{m^{2}}\right) i_{\xi} \mathfrak{J} \cdot \hat{\delta} \mathfrak{J} \\
       & +\left[ - \frac{\zeta^{2}}{\mu l^{2}} \left( \frac{3}{4} - \nu^{2} \right) + l \left| \zeta \right| \left( \frac{\alpha H_{2}}{\mu} + \frac{F_{2}}{m^{2}}\right) \right] i_{\xi} e \cdot\hat{\delta} e \\
       & -\left( \frac{\alpha H_{2}}{\mu} + \frac{F_{2}}{m^{2}}\right) \left[ i_{\xi} \mathfrak{J} \cdot \hat{\delta} \Omega + \left( i_{\xi} \Omega - \chi _{\xi}\right)\cdot \hat{\delta} \mathfrak{J} \right] \\
       & +\left[ \frac{\zeta^{2}}{\mu l^{4}} \left( 1 - \nu^{2} \right) - \frac{3\left| \zeta \right|}{2l} \left( \frac{\alpha H_{2}}{\mu} + \frac{F_{2}}{m^{2}}\right) \right] \left( i_{\xi} \mathfrak{J} \cdot \hat{\delta} e + i_{\xi}e \cdot \hat{\delta} \mathfrak{J} \right) \biggr\}.
  \end{split}
\end{equation}
where $ \mathfrak{J}^{a}_{\hspace{1.5 mm} \mu}=J^{a}J_{\mu} $ for simplicity.
One can show that the expression \eqref{98} for $\chi_{\xi}$ can be rewritten as \cite{18}
\begin{equation}\label{43}
  i_{\xi} \Omega - \chi _{\xi} = - \frac{1}{2} \varepsilon ^{a}_{\hspace{1.5 mm} bc} e^{b \mu} e^{c \nu} \nabla _{\mu} \xi _{\nu} .
\end{equation}
Also we mention that the torsion free spin-connection is given by
\begin{equation}\label{44}
  \Omega ^{a} _{ \hspace{1.5 mm}\mu} = \frac{1}{2} \varepsilon^{a b c} e _{b} ^{ \hspace{1.5 mm} \alpha} \nabla _{\mu} e_{c \alpha}.
\end{equation}
Now we take spacelike warped AdS$_{3}$ spacetime as background which can be described by the following dreibein
\begin{equation}\label{45}
  \begin{split}
     \bar{e}^{0}= & \frac{l \nu}{\sqrt{1-\nu^{2}}} dt, \\
     \bar{e}^{1}= & \frac{l}{\zeta \nu r} dr, \\
     \bar{e}^{2}= & \frac{l}{\sqrt{1-\nu^{2}}}dt+\frac{1}{2} rl\left| \zeta \right| \sqrt{1-\nu^{2}}d\phi.
  \end{split}
\end{equation}
The bar sign on the top of a quantity emphasis that the considered quantity has calculated on background. As we mentioned in section \ref{1.0}, one can take an integration from \eqref{41} over the one-parameter path on the solution space to find the conserved charge corresponds to the Killing vector field $\xi$, then
\begin{equation}\label{46}
  \begin{split}
       Q(\xi) = \frac{1}{8\pi G} \int_{\Sigma}\biggl\{& - \left( \sigma + \frac{\alpha H_{1}}{\mu} + \frac{F_{1}}{m^{2}}\right) \left[ i_{\xi} \bar{e} \cdot \Delta \Omega + \left( i_{\xi} \bar{\Omega} - \bar{\chi} _{\xi}\right)\cdot \Delta e \right]  \\
       & +\frac{1}{\mu} \left( i_{\xi} \bar{\Omega} - \bar{\chi} _{\xi}\right) \cdot \Delta \Omega + \alpha H_{2} \left( \frac{\alpha H_{2}}{\mu} + \frac{2 F_{2}}{m^{2}}\right) i_{\xi} \bar{\mathfrak{J}} \cdot \Delta \mathfrak{J} \\
       & +\left[ - \frac{\zeta^{2}}{\mu l^{2}} \left( \frac{3}{4} - \nu^{2} \right) + l \left| \zeta \right| \left( \frac{\alpha H_{2}}{\mu} + \frac{F_{2}}{m^{2}}\right) \right] i_{\xi} \bar{e} \cdot \Delta e \\
       & -\left( \frac{\alpha H_{2}}{\mu} + \frac{F_{2}}{m^{2}}\right) \left[ i_{\xi} \bar{\mathfrak{J}} \cdot \Delta \Omega + \left( i_{\xi} \bar{\Omega} - \bar{\chi} _{\xi}\right)\cdot \Delta \mathfrak{J} \right] \\
       & +\left[ \frac{\zeta^{2}}{\mu l^{4}} \left( 1 - \nu^{2} \right) - \frac{3\left| \zeta \right|}{2l} \left( \frac{\alpha H_{2}}{\mu} + \frac{F_{2}}{m^{2}}\right) \right] \left( i_{\xi} \bar{\mathfrak{J}} \cdot \Delta e + i_{\xi}\bar{e} \cdot \Delta \mathfrak{J} \right) \biggr\},
  \end{split}
\end{equation}
with $\Delta \Phi = \Phi_{(s=1)}-\Phi_{(s=0)}$, where $\Phi_{(s=1)}$ and $\Phi_{(s=0)}$ are calculated on considered spacetime solution and on the background spacetime solution, respectively\footnote{For instance, $\Delta e = e - \bar{e}$, where $e$ and $\bar{e}$ are given by \eqref{28}-\eqref{31} and \eqref{45}, respectively.}.
\section{The algebra of conserved charges}\label{7.0}
As we saw in section \ref{5.0}, the asymptotic Killing vector field is given by Eq.\eqref{32}. Let's find out conserved charge corresponds to the asymptotic Killing vector field given in Eq.\eqref{32}. To this end, we need to use \eqref{15}, \eqref{21}-\eqref{26}, \eqref{28}-\eqref{32}, \eqref{40} and \eqref{43}-\eqref{46}. After tedious calculations we find the following expression for conserved charge corresponds to the asymptotic Killing vector field \eqref{32}
\begin{equation}\label{47}
  Q(T,Y)=P(T)+L(Y),
\end{equation}
with
\begin{equation}\label{48}
  P(T)=-\frac{\left| \zeta \right|}{96 \pi} c_{U} \int_{0}^{2\pi} T(\phi) \left[ A_{rr}(\phi) +2 A_{t\phi}(\phi) \right]d \phi,
\end{equation}
\begin{equation}\label{49}
  L(Y)=\frac{\zeta ^{4} \nu ^{4}}{768 \pi}c_{V}\int_{0}^{2\pi} Y(\phi) \left[-3 A_{rr}(\phi)^{2} +4 B_{rr}(\phi) +16 B_{t\phi}(\phi) \right] d \phi ,
\end{equation}
where
\begin{equation}\label{50}
  c_{U}=\frac{3 l \left| \zeta \right| \nu^{2}}{G}\left\{ \sigma + \frac{\alpha}{\mu} \left(H_{1}+l^{2}H_{2} \right)+\frac{1}{m^{2}} \left(F_{1}+l^{2}F_{2} \right)-\frac{\left| \zeta \right|}{2\mu l}\right\},
\end{equation}
\begin{equation}\label{51}
  c_{V}=\frac{3 l }{\left| \zeta \right| \nu^{2} G}\left\{ \sigma + \frac{\alpha}{\mu} \left(H_{1}+l^{2}H_{2} \right)+\frac{1}{m^{2}} \left(F_{1}+l^{2}F_{2} \right)-\frac{\left| \zeta \right|}{2\mu l} \left( 1-2\nu^{2}\right)\right\}.
\end{equation}
It is worth to mention that $c_{U}$ and $c_{V}$ are related as
\begin{equation}\label{52}
  \zeta^{2} \nu^{4} c_{V}-c_{U}=\frac{3 \zeta^{2} \nu^{4}}{\mu G}.
\end{equation}
The algebra of conserved charges can be written as \cite{19,20}
\begin{equation}\label{53}
  \left\{ Q(\xi _{1}) , Q(\xi _{2}) \right\} = Q \left(  \left[ \xi _{1} , \xi _{2} \right] \right) + \mathcal{C} \left( \xi _{1} , \xi _{2} \right)
\end{equation}
where $\mathcal{C} \left( \xi _{1} , \xi _{2} \right)$ is central extension term. Also, the left hand side of the equation \eqref{53} can be defined by
\begin{equation}\label{54}
  \left\{ Q(\xi _{1}) , Q(\xi _{2}) \right\}= \frac{1}{2} \left( \hat{\delta} _{\xi _{2}} Q(\xi _{1}) - \hat{\delta} _{\xi _{1}} Q(\xi _{2}) \right).
\end{equation}
Therefore one can obtain the central extension term by using the following formula
\begin{equation}\label{55}
  \mathcal{C} \left( \xi _{1} , \xi _{2} \right)= \frac{1}{2} \left( \hat{\delta} _{\xi _{2}} Q(\xi _{1}) - \hat{\delta} _{\xi _{1}} Q(\xi _{2}) \right) - Q \left(  \left[ \xi _{1} , \xi _{2} \right] \right).
\end{equation}
Thus, one can use Eq.\eqref{33}, Eq.\eqref{36}, Eq.\eqref{37} and Eq.\eqref{47} to find that
\begin{equation}\label{56}
  \begin{split}
     \left\{ Q(T_{1},Y_{1}) , Q(T_{2},Y_{2}) \right\} =& Q(T_{12},Y_{12}) \\
       & +\frac{\left| \zeta \right|}{192 \pi} c_{U} \int_{0}^{2\pi} T_{12}(\phi) \left[ A_{rr}(\phi) +2 A_{t\phi}(\phi) \right]d \phi\\
       & -\frac{1}{48 \pi} c_{U} \int_{0}^{2\pi} \left( T_{1} \partial_{\phi} T_{2}-T_{2} \partial_{\phi} T_{1}\right)d \phi \\
       & -\frac{1}{48 \pi}c_{V} \int_{0}^{2\pi} \left( Y_{1} \partial_{\phi}^{3} Y_{2}-Y_{2} \partial_{\phi}^{3} Y_{1}\right)d \phi.
  \end{split}
\end{equation}
We introduce Fourier modes as
\begin{equation}\label{57}
  \begin{split}
      P_{m} = & Q(e^{im\phi},0)=P(e^{im\phi}), \\
      L_{m} = & Q(0,e^{im\phi})=L(e^{im\phi}),
  \end{split}
\end{equation}
so one can read off the algebra of conserved charges as follows:
\begin{equation}\label{58}
  \begin{split}
     i \left\{ P_{m} , P_{n} \right\} =& -\frac{c_{U}}{12} m \delta_{m+n,0}  \\
     i \left\{ L_{m} , P_{n} \right\} =& -n P_{m+n} - \frac{\left| \zeta \right| c_{U}}{192 \pi}n \int_{0}^{2\pi} e^{i(m+n)\phi} \left[ A_{rr}(\phi) +2 A_{t\phi}(\phi) \right]d \phi \\
     i \left\{ L_{m} , L_{n} \right\} =& (m-n) L_{m+n} +\frac{c_{V}}{12}m^{3} \delta_{m+n,0}.
  \end{split}
\end{equation}
Now we consider warped black hole solution as an example. For warped black hole \eqref{12} with\eqref{13}, we have
\begin{equation}\label{59}
  \begin{split}
       & A_{rr}=r_{+}+r_{-}, \hspace{0.7 cm} A_{t\phi}=\nu \sqrt{r_{+}r_{-}}, \\
       & B_{rr}=r_{+}^{2}+r_{-}^{2}+r_{+}r_{-}, \hspace{0.7 cm} B_{t\phi}=0.
  \end{split}
\end{equation}
In this case, equations \eqref{58} will be reduce to
\begin{equation}\label{60}
  \begin{split}
     i \left\{ P_{m} , P_{n} \right\} =& -\frac{c_{U}}{12} m \delta_{m+n,0}  \\
     i \left\{ L_{m} , P_{n} \right\} =& -n P_{m+n} - \frac{\left| \zeta \right| c_{U}}{96}n \left( r_{+}+r_{-}+2\nu \sqrt{r_{+}r_{-}} \right) \delta_{m+n,0}\\
     i \left\{ L_{m} , L_{n} \right\} =& (m-n) L_{m+n} +\frac{c_{V}}{12}m^{3} \delta_{m+n,0}.
  \end{split}
\end{equation}
Now we set $ \hat{P} _{m} \equiv P_{m}$ and $ \hat{L} _{m} \equiv L_{m}$, also we replace Dirac brackets by commutators $i \{ , \} \rightarrow [,]$, therefore we can rewritten equations \eqref{58} as following
\begin{equation}\label{61}
  \begin{split}
      \left[ \hat{P}_{m} , \hat{P}_{n} \right] =& -\frac{c_{U}}{12} m \delta_{m+n,0}  \\
      \left[ \hat{L}_{m} , \hat{P}_{n} \right] =& -n \hat{P}_{m+n} +\frac{n}{2}p_{0}\delta_{m+n,0} \\
      \left[ \hat{L}_{m} , \hat{L}_{n} \right] =& (m-n) \hat{L}_{m+n} +\frac{c_{V}}{12}m^{3} \delta_{m+n,0}
  \end{split}
\end{equation}
where $p_{0}$ is the zero mode eigenvalue\footnote{ Here, it is reasonable to use the terminology "eigenvalues"
 because we improve $P_{m}$ and $L_{m}$ to be operators $ \hat{P} _{m}$ and $ \hat{L} _{m}$, respectively,
 and we denote eigenvalues of them by $p_{m}$ and $l_{m}$. Such a terminology have been used by
Ba$\tilde{\text{n}}$ados \cite{29}, Strominger \cite{13'}, Carlip \cite{30} and so on.} of $\hat{P}_{m} $. From Eq.\eqref{57} and using Eq.\eqref{48}, Eq.\eqref{49} and Eq.\eqref{59}, one can easily read off the eigenvalues of $\hat{P}_{m} $ and $\hat{L}_{m} $ as
\begin{equation}\label{62}
  p_{m}=- \frac{\left| \zeta \right| c_{U}}{48} \left( r_{+}+r_{-}+2\nu \sqrt{r_{+}r_{-}} \right)
\delta_{m,0},
\end{equation}
\begin{equation}\label{63}
  l_{m}=\frac{\zeta^{4} \nu^{4}c_{V}}{384} \left( r_{+}-r_{-}\right)^{2} \delta_{m,0},
\end{equation}
respectively. It is clear now that the algebra of asymptotic conserved charges is given as the semi direct product of the Virasoro algebra with $U(1)$ current algebra, with central charges $c_{V}$ and $c_{U}$.
\section{Mass, angular momentum and entropy of warped black hole solution of GMMG}\label{8.0}
It is clear that the algebra of conserved charges \eqref{61} does not describe the conformal symmetry \cite{21}. However, one can use a particular Sugawara construction \cite{22} to reconstruct the conformal algebra. Such a method has been used in the topologically massive gravity model \cite{23}, also such study on spacelike warped AdS$_{3}$ black hole solutions of NMG has been done in \cite{26}. Therefore, we introduce two new operators as follows
\begin{equation}\label{64}
  \hat{L}^{+}_{m}=\frac{im}{\left| \zeta \right| \nu^{2}} \hat{P}_{-m} + \frac{6}{c_{U}} \hat{K}_{-m},
\end{equation}
\begin{equation}\label{65}
  \hat{L}^{-}_{m}=\hat{L}_{m} - \frac{6p_{0}}{c_{U}} \hat{P}_{m} + \frac{6}{c_{U}} \hat{K}_{m},
\end{equation}
where
\begin{equation}\label{66}
  \hat{K}_{m}=\sum_{q \in \mathbb{Z}} \hat{P}_{m+q} \hat{P}_{-q}.
\end{equation}
One can show that the algebra among $\hat{L}^{\pm}_{m}$ is given as
\begin{equation}\label{67}
\begin{split}
   \left[ \hat{L}^{+}_{m} , \hat{L}^{+}_{n} \right] = & (m-n) \hat{L}^{+}_{m+n} +\frac{c_{+}}{12}m^{3} \delta_{m+n,0} \\
   \left[ \hat{L}^{-}_{m} , \hat{L}^{-}_{n} \right] = & (m-n) \hat{L}^{-}_{m+n} +\frac{c_{-}}{12}m^{3} \delta_{m+n,0}+\frac{3p_{0}^{2}}{c_{U}}m \delta_{m+n,0} \\
   \left[ \hat{L}^{+}_{m} , \hat{L}^{-}_{n} \right] = &0
\end{split}
\end{equation}
where
\begin{equation}\label{68}
  c_{+}= \frac{c_{U}}{\zeta^{2} \nu^{4}}, \hspace{0.7 cm} c_{-}=c_{V},
\end{equation}
In this way, Eq.\eqref{52} can be rewritten as
\begin{equation}\label{69}
  c_{-}-c_{+}=\frac{3}{\mu G}.
\end{equation}
It is easy to see, from Eq.\eqref{64} and Eq.\eqref{65}, that the values of $\hat{L}^{\pm}_{0}$ are given as
\begin{equation}\label{70}
  l^{+}_{0}=\frac{6p_{0}^{2}}{c_{U}}, \hspace{0.7 cm} l^{-}_{0}=l_{0}.
\end{equation}
Now we can calculate the black hole entropy via Cardy's formula \cite{24,25},
\begin{equation}\label{71}
  \mathcal{S_{CFT}}=2\pi \sqrt{\frac{c_{+}l^{+}_{0}}{6}}+2\pi \sqrt{\frac{c_{-}l^{-}_{0}}{6}}
\end{equation}
and thus we find that
\begin{equation}\label{72}
\begin{split}
   \mathcal{S_{CFT}}= & \frac{\pi l \left| \zeta \right|}{4G} \biggl\{ \left[ \sigma + \frac{\alpha}{\mu} \left(H_{1}+l^{2}H_{2} \right)+\frac{1}{m^{2}} \left(F_{1}+l^{2}F_{2} \right)-\frac{\left| \zeta \right|}{2\mu l} \right] \left( r_{+} + \nu \sqrt{r_{+} r_{-}} \right)\\
     & + \frac{\left| \zeta \right| \nu^{2}}{2 \mu l} \left( r_{+}-r_{-}\right) \biggr\}.
\end{split}
\end{equation}
Mass and angular momentum can also be obtained by $\mathcal{M}=-p_{0}$ and $\mathcal{J}=-\left( l^{+}_{0}-l^{-}_{0}\right)$, respectively \cite{23}. Thus, we have
\begin{equation}\label{73}
  \mathcal{M}=\frac{\left| \zeta \right| c_{U}}{48} \left( r_{+}+r_{-}+2\nu \sqrt{r_{+}r_{-}} \right),
\end{equation}
\begin{equation}\label{74}
  \mathcal{J}=-\frac{\zeta^{2}}{384} \left\{ c_{U} \left( r_{+}+r_{-}+2\nu \sqrt{r_{+}r_{-}} \right)^{2} - \zeta^{2} \nu^{4} c_{V} \left( r_{+}-r_{-}\right)^{2} \right\}.
\end{equation}
As we know, angular velocity and surface gravity of horizon are given by
\begin{equation}\label{75}
  \Omega _{H} =  - N^{\phi}(r_{+})=-\frac{2}{\left| \zeta \right| \left( r_{+} + \nu \sqrt{r_{+} r_{-}} \right)},
\end{equation}
\begin{equation}\label{76}
  \kappa_{H}= \left[ -\frac{1}{2} \nabla^{\mu} k^{\nu} \nabla_{\mu} k_{\nu} \right]^{\frac{1}{2}}_{r=r_{+}}=\frac{\left| \zeta \right| \nu^{2}\left( r_{+}-r_{-} \right)}{2\left( r_{+} + \nu \sqrt{r_{+} r_{-}} \right)},
\end{equation}
respectively, where $k=\partial_{t}+\Omega_{H} \partial_{\phi}$ is the horizon-generating Killing vector field. One can verify that mass \eqref{73}, angular momentum \eqref{74} and entropy \eqref{72} of considered black hole satisfy the first law of black hole mechanics,
\begin{equation}\label{77}
  \delta \mathcal{M}=T_{H} \delta \mathcal{S} + \Omega_{H} \delta \mathcal{J},
\end{equation}
where $T_{H}=\kappa_{H}/2\pi$ is the Hawking temperature.\\
Let us conclude this section with the following important remark.
 We know that the $SL(2,\mathbb{R})\times SL(2,\mathbb{R}$ isometry group of AdS$_{3}$ space reduce to
 the $SL(2,\mathbb{R})\times U(1)$ in warped AdS$_{3}$ space due to the presence of warping parameter.
 So one expect that the asymptotic symmetry of warped AdS$_{3}$ space also differ from full conformal symmetry.
 So the dual theory of warped AdS$_{3}$ space instead a 2-dimensional CFT, is warped CFT (WCFT), which exhibits
 partial conformal symmetry. In another term, for this case the conformal symmetry is not present, actually not even
 Lorentz symmetry.
 Let us define $\tilde{P}_{m} = \left| \zeta \right|^{-1} \nu^{-2} \hat{P}_{m}$.
 By that definition the algebra \eqref{61} becomes
\begin{equation}\label{81}
  \begin{split}
      \left[ \tilde{P}_{m} , \tilde{P}_{n} \right] =& -\frac{\tilde{c}_{U}}{12} m \delta_{m+n,0}  \\
      \left[ \hat{L}_{m} , \tilde{P}_{n} \right] =& -n \tilde{P}_{m+n} +\frac{n}{2}\tilde{p}_{0}\delta_{m+n,0} \\
      \left[ \hat{L}_{m} , \hat{L}_{n} \right] =& (m-n) \hat{L}_{m+n} +\frac{c_{V}}{12}m^{3} \delta_{m+n,0}
  \end{split}
\end{equation}
where $\tilde{c}_{U}=\zeta^{-2} \nu^{-4} c_{U}$ and $ \tilde{p}_{m} = \left| \zeta \right|^{-1} \nu^{-2} p_{m}$.
 The algebra \eqref{81} is the semi direct product of the Virasoro algebra with $U(1)$ current algebra, with central
 charges $c_{V}$ and $\tilde{c}_{U}$. This is exactly the symmetry of warped CFT, so the dual theory of warped
black hole solution of GMMG is a WCFT. This conformal field theory in the so-called quadratic ensemble has
 an entropy in the Cardy regime that looks like usual Cardy's formula for entropy in a CFT.
This warped Cardy's formula which has been introduced by Detournay et. al \cite{27} present the asymptotic
 growth of state in the WCFT. Now, we can use the warped Cardy formula which is given by \cite{27}
\begin{equation}\label{82}
  \mathcal{S}_{\text{WCFT}}=\frac{24\pi}{\tilde{c}_{U}}\tilde{p}_{0}^{(vac)}\tilde{p}_{0}+
4\pi \sqrt{-l_{0}^{(vac)}l_{0}},
\end{equation}
where $ \tilde{p}_{0}^{(vac)}$ and $ l_{0}^{(vac)}$ correspond to the minimum values of
 $ \tilde{p}_{0}$ and $ l_{0}$, i.e. the value of the vacuum geometry. Since vacuum corresponds to $r_{\pm}=0$,
 so from Eq.\eqref{81}, one can read off
\begin{equation}\label{83}
  \tilde{p}_{0}^{(vac)}= -\frac{\tilde{c}_{U}}{24}, \hspace{0.7 cm} l_{0}^{(vac)}= -\frac{c_{V}}{24}.
\end{equation}
By substituting Eq.\eqref{62}, Eq.\eqref{63} and Eq.\eqref{83} into
 Eq.\eqref{82}, we find that $\mathcal{S}_{\text{WCFT}}=\mathcal{S}_{\text{CFT}} $.
\section{Warped black hole entropy from gravitational point of view}\label{9.0}
The gravitational black hole entropy can be obtained by Eq.\eqref{5} in CSLTG. Now we want to simplify Eq.\eqref{5} for GMMG model which is an example of CSLTG. Also, we are interested to find the entropy of Warped black hole entropy \eqref{12}, with Eq.\eqref{13}, as a solution of GMMG model. One can use Eq.\eqref{6} and Eq.\eqref{20} to show that, for the considered case, the gravitational black hole entropy is simplified as
\begin{equation}\label{78}
\begin{split}
    \mathcal{S}=- \frac{1}{4G} \int_{\text{Horizon}} \frac{d\phi}{\sqrt{g_{\phi \phi}}} \biggl\{ & -\left(\sigma + \frac{\alpha H_{1}}{\mu} + \frac{F_{1}}{m^{2}} \right) g_{\phi \phi} + \frac{1}{\mu} \Omega_{\phi \phi} \\
     & - \left( \frac{\alpha H_{2}}{\mu} + \frac{F_{2}}{m^{2}} \right) J_{\phi} J_{\phi} \biggr\}.
\end{split}
\end{equation}
Also, for warped black hole solution \eqref{12} with \eqref{13}, we have
\begin{equation}\label{79}
  \begin{split}
      g_{\phi \phi} \mid _{r=r_{+}} = & \frac{1}{4} l^{2} \zeta^{2} \left( r_{+} + \nu \sqrt{r_{+}r_{-}}\right)^{2}, \\
      \Omega_{\phi \phi} \mid _{r=r_{+}}=& -\frac{1}{4} \zeta^{2} \nu^{2} \sqrt{g_{\phi \phi}} \mid _{r=r_{+}} \left( r_{+}-r_{-} \right) + \frac{\left| \zeta \right|}{2 l} g_{\phi \phi}\mid _{r=r_{+}}, \\
      \left( J_{\phi} J_{\phi} \right) \mid _{r=r_{+}}= & l^{2} g_{\phi \phi}\mid _{r=r_{+}}.
  \end{split}
\end{equation}
By substituting Eq.\eqref{79} into Eq.\eqref{80}, we find the gravitational entropy of warped black hole as
\begin{equation}\label{80}
\begin{split}
   \mathcal{S}= & \frac{\pi l \left| \zeta \right|}{4G} \biggl\{ \left[ \sigma + \frac{\alpha}{\mu} \left(H_{1}+l^{2}H_{2} \right)+\frac{1}{m^{2}} \left(F_{1}+l^{2}F_{2} \right)-\frac{\left| \zeta \right|}{2\mu l} \right] \left( r_{+} + \nu \sqrt{r_{+} r_{-}} \right)\\
     & + \frac{\left| \zeta \right| \nu^{2}}{2 \mu l} \left( r_{+}-r_{-}\right) \biggr\}.
\end{split}
\end{equation}
This result exactly coincide with what we obtained in the previous section \eqref{72} via Cardy's formula.
\section{Conclusion}
The Chern-Simons-like theories of gravity describe by the Lagrangian \eqref{1}. In such theories, the quasi-local conserved charge perturbation associated with a field-dependent vector field $\xi$ is given by Eq.\eqref{3}. By integrating from Eq.\eqref{3} over the one-parameter path on the solution space, one finds the quasi-local conserved charge \eqref{4} associated with a field-dependent vector field $\xi$. The generalized minimal massive gravity (GMMG), is an example of the Chern-Simons-like theories of gravity, which is described by the equations of motion \eqref{7}-\eqref{10}. Warped black hole spacetime is given by the metric \eqref{12} with \eqref{13}. In section \ref{4.0}, we showed that the warped black hole spacetime solves equations of motion of GMMG. In section \ref{5.0}, we have introduced boundary conditions \eqref{27}, which describes asymptotically spacelike warped AdS$_{3}$ spacetimes at spatial infinity. The variation generated by the Killing vector field \eqref{32} preserves the considered boundary conditions. The asymptotic Killing vectors \eqref{32} are closed in the Lie bracket ( see Eq.\eqref{33}). By introducing Fourier modes, we saw that the algebra among the asymptotic Killing vectors is the semi direct product of the Witt algebra with the $U(1)$ current algebra. Under the action of a generic asymptotic symmetry generator $\xi$ spanned by \eqref{32}, the dynamical fields appeared in the metric \eqref{27} transform as \eqref{36} and \eqref{37}. Also,
we have shown that the metric \eqref{27} solves equations of motion of GMMG asymptotically when equations \eqref{21}-\eqref{26} and \eqref{40} are satisfied. We found that the conserved charge, corresponds to the asymptotic Killing vector field \eqref{32}, is given by Eq.\eqref{47}. Also, in section \ref{7.0}, we showed that the algebra among the conserved charges is given by \eqref{58}. In Eq.\eqref{58}, we set $ \hat{P} _{m} \equiv P_{m}$ and $ \hat{L} _{m} \equiv L_{m}$, also we replaced the Dirac brackets by commutators $i \{ , \} \rightarrow [,]$, then we obtained Eq.\eqref{61}. The algebra of asymptotic conserved charges is given as the semi direct product of the Virasoro algebra with $U(1)$ current algebra, with central charges $c_{V}$ and $c_{U}$. The algebra of the conserved charges \eqref{61} does not describe the conformal symmetry. Therefore, we used a particular Sugawara construction to reconstruct the conformal algebra ( see Eqs.\eqref{64}-\eqref{68}). Thus, we are allowed to use the Cardy's formula to calculate the entropy of the warped black hole and then we have obtained that the warped black hole entropy is given by Eq.\eqref{72}. Also, we showed that mass and angular momentum of the warped black hole are given by Eq.\eqref{73} and Eq.\eqref{74}, respectively. The obtained mass, angular momentum and entropy satisfy the first law of black hole mechanics. As we mentioned in section \ref{9.0}, the gravitational black hole entropy can be obtained by Eq.\eqref{5} in CSLTG. Hence, in section \ref{9.0}, we have shown that the gravitational entropy of warped black hole in GMMG model is given by Eq.\eqref{80}. By comparing Eq.\eqref{72}, Eq.\eqref{82}  and Eq.\eqref{80}, we see that the gravitational entropy of the warped black hole exactly coincide with what we obtained via Cardy's formula and the warped version of this formula. Although we could showed that by Sugawara construction one can obtain an asymptotic symmetry for the warped AdS$_{3}$ black hole solution of GMMG that coincides with the conformal symmetry, described by two independent Virasoro algebra. But we do not claimed that the dual theory of such gravity solutions of GMMG is CFT. It has been argued by  El-Showk, and Guica \cite{28} that warped AdS$_{3}$  spaces could be dual to certain types of non-local CFT. Since the $SL(2,\mathbb{R})\times SL(2,\mathbb{R}$ isometry group of AdS$_{3}$ space reduce to the $SL(2,\mathbb{R})\times U(1)$ in warped AdS$_{3}$ space due to the presence of warping parameter. So one expect that the asymptotic symmetry of warped AdS$_{3}$ space also differ from full conformal symmetry. So the dual theory of warped AdS$_{3}$ space instead a 2-dimensional CFT, is warped CFT (WCFT), which exhibits partial conformal symmetry.
\section{Acknowledgments}
M. R. Setare thank Stephane Detournay  and Tom Hartman for helpful comments and discussions.

\end{document}